\documentclass[english,aps,amsmath,amssymb,superscriptaddress,twocolumn,prb]{revtex4}
\usepackage{graphicx}
\usepackage{dcolumn}
\usepackage{bm}
\usepackage{subfigure}
\usepackage{float}
\usepackage{xcolor}

\begin{document}

\title{A Standard Basis Operator Equation of Motion Impurity Solver for Dynamical Mean Field Theory}

\author{Hengyue Li}
\affiliation{Department of Physics, Renmin University of China, 100872 Beijing, China}
\affiliation{Beijing Key Laboratory of Opto-electronic Functional Materials and Micro-nano Devices (Renmin University of China)}

\author{Ning-Hua Tong}
\email{nhtong@ruc.edu.cn}
\affiliation{Department of Physics, Renmin University of China, 100872 Beijing, China}
\affiliation{Beijing Key Laboratory of Opto-electronic Functional Materials and Micro-nano Devices (Renmin University of China)}
\date{\today}

\begin{abstract}
We present an efficient impurity solver for the dynamical mean-field theory (DMFT). It is based on the separation of bath degrees of freedom into the low energy and the high energy parts. The former is solved exactly using exact diagonalization and the latter is treated approximately using Green's function equation of motion decoupling approximation. The two parts are combined coherently under the standard basis operator formalism. The impurity solver is applied to the Anderson impurity model and, combined with DMFT, to the one-band Hubbard model. Qualitative agreement is found with other well established methods. Some promising features and possible improvements of the present solver are discussed.
\end{abstract}


\maketitle
\section{Introduction}

In the past two decades, the dynamical mean field theory (DMFT)
 has been established as a powerful method for studying strongly correlated electron systems.\cite{Vollhardt1,Georges1} In DMFT, a quantum lattice model for interacting electrons is mapped onto an effective single impurity model with a self-consistently determined bath. To study the multi-band lattice Hamiltonian or systems with strong spatial fluctuations such as two dimensional models, one needs to map the original Hamiltonian onto effective multi-band impurity models,\cite{Held1,Liebsch1,Werner1,Georges2} or onto a cluster Hamiltonian with many impurities.\cite{Kotliar1,Hettler1} For DMFT and its extensions, the core is the solution of the impurity or cluster model for Green's functions, or equivalently, for the self-energies.\cite{Maier1} The efficiency of solving the quantum impurity or cluster models is essential for the applicability of DMFT and its extensions.

Due to its obvious importance, fast and accurate method for solving impurity and cluster models has been the goal of intensive research activities since the development of DMFT. In this area, various quantum Monte Carlo (QMC) methods, such as Hirsch-Fye algorithm\cite{Hirsch1,Bluemer1} and strong\cite{Werner2}/weak-coupling\cite{Rubtsov1} continuous time QMC are among the most powerful and widely used methods. They are numerically exact and can treat multi-band impurity models and the multi-impurity cluster models.\cite{Gull1} However, these QMC methods cannot reach zero temperature and the dynamical correlations on real frequency axis are produced through numerical analytical continuation, which introduces additional errors.\cite{Jarrell1} Numerical renormalization group method\cite{Wilson1} can produce accurate low energy properties for single impurity model, but the computational cost increase exponentially with the number of orbitals or impurities.\cite{Bulla1} The full exact diagonalization (ED)\cite{Caffarel1} and the Lanczos method\cite{Si1} are relatively fast and easy to implement. However, at present, the total number of sites of the impurity/cluster model is constrained to about 9 for ED and 15 for Lanczos.\cite{Liebsch2} This limits the energy resolution in the resulting spectral function.

Besides these more established methods, there have been attempts to develop new impurity/cluster solvers that are focused on the efficiency  for complex impurity models, allowing for approximations. Among them are the weak coupling\cite{Georges3} or strong coupling\cite{Dai1} perturbation theory, fluctuation-exchange approximation,\cite{Aryanpour1} non-crossing approximation,\cite{Bickers1} Gutzwiller approximation,\cite{Zhuang1} double time Green's function (GF) equation of motion (EOM) method,\cite{Gros1,Jeschke1,Feng1} superperturbation method,\cite{Hafermann1} extended recursion in operator space,\cite{Julien1} and the distribution ED method,\cite{Granath1} {\it etc}. Recently, two methods appear to be able to go beyond the limitations in QMC, NRG and ordinary ED. One is the ED algorithm based on natural orbitals,\cite{He1} and the other is the hierarchical equation of motion method.\cite{Li1} These two methods seem promising in DMFT applications, but their implementations are complicated and the computational costs are relatively high.

It is the purpose of this paper to present a new impurity solver which is approximate but can be improved systematically. In this new method, the bath degrees of freedom are split into the low energy and the high energy parts. The influence of the low energy bath to the impurity is treated exactly using the traditional ED method. While the influence of the high energy bath is treated afterwards using an approximate GF EOM truncation scheme, that is, the alloy analogy approximation (AAA).\cite{Hubbard1} This idea is realized by using the standard basis operator (SBO) and its GF EOM method,\cite{Haley1} which is originally designed for extending the random phase approximation in the GF EOM formalism to the spin-$1$ Heisenberg model. Here we use it to derive a non-perturbative approximation on top of ED of a small system containing the impurity and $n_{s}$ bath sites. We hope that this method can combine the advantage of ED, that is, high speed and being systematic, with that of GF EOM approach, that is, being flexible and giving continuous spectral function. Our benchmark calculation shows that this method is fast and can produce qualitatively correct spectral function using one bath site $n_{s}=1$. Therefore, this method is suitable for DMFT study of more complicated models of strongly correlated electrons, expecting reasonable results in a limited calculation time.

In the rest part of this paper, we first present the formalism in Section II. The results of benchmarking calculations are presented and compared with ED and NRG results in Section III. In Section IV, we discuss the method and its possible extensions and applications. In Section V, a brief summary is given.

\section{Formalism}

In this part, we derive the formalism of the standard basis operator GF EOM method. For simplicity, we consider the single band Hubbard model on a Bethe lattice in the infinite coordination limit. The DMFT equation for this system can be obtained analytically and has been solved by many other established methods. It is an ideal system for  benchmarking a new solver.

The Hamiltonian of the single band Hubbard model on Bethe lattice reads
\begin{equation} \label{eq:1}
  H=-t \sum_{\langle i,j \rangle \sigma} d_{i \sigma}^{\dagger} d_{j\sigma} + \sum_{i} U n_{i \uparrow} n_{i \downarrow} -\mu \sum_{i, \sigma} n_{i \sigma}.
\end{equation}
Here, $d_{i \sigma}^{\dagger}$ and $d_{i \sigma}$ are the creation and the annihilation operators of a spin-$\sigma$ electron on site $i$, respectively. The summation $\sum_{\langle i,j \rangle}$ is for the nearest neighbour sites on a Bethe lattice with coordination number $z$. $U$ is the on-site repulsion energy and $\mu$ is the chemical potential. In DMFT, this lattice Hamiltonian is mapped onto an effective Anderson impurity model of the form
\begin{eqnarray} \label{eq:2}
   H_{Aim} =&& \sum_{k \sigma} \epsilon_{k} c_{k \sigma}^{\dagger} c_{k \sigma} + \sum_{k \sigma} V_k \left(c_{k \sigma}^{\dagger} d_{\sigma} + d_{\sigma}^{\dagger} c_{k \sigma} \right)    \nonumber \\
   && +U n_{\uparrow}n_{\downarrow} - \mu \sum_{\sigma} n_{\sigma}.
\end{eqnarray}
Here $c_{k \sigma}$ is the annihilation operator of the bath electron. The hybridization function is defined as
\begin{equation} \label{eq:3}
   \Delta(\omega) \equiv \sum_{k} V_{k}^{2} \delta(\omega - \epsilon_k),
\end{equation}
with its counter part on the Matsubara frequency axis,
\begin{equation} \label{eq:4}
   \Gamma(i \omega_n) \equiv \sum_{k} \frac{V_{k}^{2}}{ i \omega_n - \epsilon_k}.
\end{equation}
$\Delta(\omega)$ and $\Gamma(i\omega_n)$ are connected by the analytical continuation and the Hilbert transformation as below,
\begin{eqnarray} \label{eq:5}
 &&  \Delta(\omega) = - \frac{1}{\pi} \text{Im} \Gamma\left(i\omega_n \rightarrow \omega + i \eta \right), \nonumber \\
 &&  \Gamma(i\omega_n) = \int_{-\infty}^{\infty}  \frac{\Delta(\epsilon)}{i\omega_n - \epsilon} d\epsilon.
\end{eqnarray}
We split the continuous bath degrees of freedom into two parts, a discrete part with $1 \le k \le n_{s}$ and a continuous part with $k \ge n_{s}+1$. Here $n_{s}$ is the number of discrete bath sites which will be treated by ED. Their respective hybridization functions are
\begin{eqnarray} \label{eq:6}
  \Delta_{1}(\omega) &=& \sum_{k=1}^{n_{s}} V_{k}^2 \delta(\omega - \epsilon_k), \nonumber \\
  \Delta_{2}(\omega) &=& \Delta(\omega) - \Delta_1(\omega).
 \end{eqnarray}
Correspondingly, $\Gamma_{1}(i \omega_n) = \sum_{k=1}^{n_{s}} V_{k}^{2}/ \left(i\omega_n - \epsilon_k \right)$ and $\Gamma_{2}(i\omega_n) = \Gamma(i\omega_n) - \Gamma_{1}(i\omega_n)$.

Given a full continuous hybridization function $\Delta(\omega)$, we first identify $n_{s}$ discrete bath modes to define $\Delta_1(\omega)$ according to some importance criteria. The impurity problem composed of the impurity site and these $n_{s}$ bath sites are diagonalized first. The influence of the residual continuous bath $\Delta_2(\omega)$ is regarded as a modification to the ED result and is treated by approximate method. This idea has been proposed by Hafermann {\it et al.}.\cite{Hafermann1} In their work, $\{V_k, \epsilon_k \} (k=1,2,...,n_{s})$ are obtained by a weighted fitting of the original hybridization function $\Gamma(i \omega_n)$ by the $n_s$-bath-site hybridization function $\Gamma_1(i\omega_n)$. The influence of the residual hybridization $\Gamma_2(i\omega_n)$ is treated perturbatively, using skeleton diagram expansion up to third order in the interacting vertex in dual space. Although this approach produces rather accurate GF's compared to QMC, the perturbative treatment of $\Gamma_2(i\omega_n)$ needs the two-particle GF $\chi_{1234}$ obtained from ED calculation as an input. For multi-band or multi-site impurity problems, the calculation and storage of $\chi_{1234}$ is expensive. This could hinder the practical applications of this approach. In this work, we use GF EOM with truncation approximation, which involves only $2$-time GF's and thus avoids the direct calculation of $4$-time GFs. For the parameterization of the discrete bath modes, we fit the low energy part of the full hybridization, which is important for the accurate description of the Kondo resonance in the metallic phase (see below).

In accordance with the splitting of bath, $H_{Aim}$ is also split into two parts,
\begin{eqnarray} \label{eq:7}
  H_{Aim} && = H_0 + H_1,   \nonumber \\
  H_0 = && \sum_{k=1}^{n_{s}} \sum_{\sigma} \epsilon_{k} c_{k \sigma}^{\dagger} c_{k \sigma} + \sum_{k=1}^{n_{s}}\sum_{\sigma} V_k \left(c_{k \sigma}^{\dagger} d_{\sigma} + d_{\sigma}^{\dagger} c_{k \sigma} \right)    \nonumber \\
        && + U n_{\uparrow}n_{\downarrow} - \mu \sum_{\sigma} n_{\sigma} ,
                                             \nonumber  \\
   H_1 =&& \sum_{k \ge n_{s}+1, \sigma} \epsilon_{k} c_{k \sigma}^{\dagger} c_{k \sigma} + \sum_{k \ge n_{s}+1, \sigma} V_k \left(c_{k \sigma}^{\dagger} d_{\sigma} + d_{\sigma}^{\dagger} c_{k \sigma} \right) .     \nonumber \\
  &&
\end{eqnarray}
To integrate ED and GF EOM in a coherent way, we use the SBO formalism proposed by Haley.\cite{Haley1,Haley2} Suppose $H_0$ has been diagonalized and let $E_{\mu}$ and $|\mu \rangle$ be the eigen energy and eigen state of $H_0$, respectively. That is, $H_0 | \mu \rangle = E_{\mu} | \mu \rangle$. SBO is defined as an projection operator in the Hilbert space of $H_0$ in the $| \mu \rangle$ basis, $A_{\mu \nu} \equiv |\mu \rangle \langle \nu |$. Suppose that the Hilbert space of $H_0$ is $D$ dimensional, the total number of SBO's is $D^2$. The algebraic relations among these $A_{\mu \nu}$'s are
\begin{eqnarray} \label{eq:8}
  &&   \left( A_{\mu \nu} \right)^{\dagger} = A_{\nu \mu}   \nonumber \\
  &&   A_{\mu \nu} A_{\alpha \beta} = \delta_{\nu \alpha} A_{\mu \beta}  \nonumber \\
  &&   \sum_{\mu} A_{\mu \mu} = 1.
\end{eqnarray}
Clearly, the $D^2$ SBO's defined above form a complete and linear-independent basis set in the operator space built on $H_0$'s Hilbert space. Any operator in this space can be expanded by $A_{\mu\nu}$'s. Especially we have
\begin{eqnarray}  \label{eq:9}
  &&   H_0 = \sum_{\mu} E_{\mu} A_{\mu \mu},   \nonumber \\
  &&   d_{\sigma}  = \sum_{\mu\nu} f_{\mu\nu}^{\sigma} A_{\mu \nu}^{\sigma}.
\end{eqnarray}
Here the coefficients $f_{\mu \nu}^{\sigma} = \langle \mu | d_{\sigma}|\nu \rangle$. Usually the coefficients $\{ f_{\mu \nu}^{\sigma} \}$ form a sparse matrix and the sparseness can be used to simplify calculations. The superscript $\sigma$ in $A_{\mu\nu}^{\sigma}$ is used to remind us that this SBO appears in the expansion of $d_{\sigma}$. For the particle- and spin- conserving Hamiltonian $H_0$, the excitation operators $A_{\mu\nu}$ can be classified by its quantum numbers $\delta N$ and $\delta S_z$ defined as $ \left[ \hat{N}, A_{\mu\nu} \right]= \delta N A_{\mu\nu}$ and $ \left[ \hat{S_z}, A_{\mu\nu} \right] = \delta S_z A_{\mu\nu}$. Here $\hat{N}$ and $\hat{S_z}$ are the total electron number operator and the z-component of total spin operator, respectively. The $A_{\mu\nu}^{\sigma}$'s in Eq.(\ref{eq:9}) are SBO's with quantum numbers $\left( \delta N = -1, \delta S_z = -\sigma/2 \right)$. Here $\sigma=1$ denotes spin up and $\sigma=-1$ spin down. In the following, we will use $A_{\mu \nu}^{\sigma}$ as defined here, and $A_{\mu \nu}$ for general SBO's with unspecified quantum numbers.

In terms of SBO's defined above, $H_{Aim}$ can be written as
\begin{eqnarray} \label{eq:10}
  H_{Aim} = &&  \sum_{\mu\nu} E_{\mu} A_{\mu\nu} + \sum_{k=n_{s}+1}^{\infty}\sum_{\sigma} \epsilon_k c_{k\sigma}^{\dagger} c_{k \sigma}    \nonumber \\
  &+& \sum_{k=n_{s}+1}^{\infty} V_{k} \left( c_{k\sigma}^{\dagger} d_{\sigma}
       +  d_{\sigma}^{\dagger} c_{k\sigma} \right) ,
\end{eqnarray}
where $d_{\sigma} = \sum_{\alpha \beta} f_{\alpha \beta}^{\sigma} A_{\alpha \beta}^{\sigma}$ and $d_{\sigma}^{\dagger} = \sum_{\gamma \delta} f_{\gamma \delta}^{\sigma \ast} A_{\gamma \delta}^{\sigma \dagger} $. Written in this form, $H_{Aim}$ can be regarded as a new problem, describing a complicated impurity coupled to a reduced bath with the effective hybridization function $\Gamma_2(i \omega_n)$.

The impurity Green's function in Zubarev's symbol is $\langle \langle d_{\sigma}| d_{\sigma}^{\dagger}  \rangle \rangle_{\omega}$. It is expressed by the GF of SBO's as
\begin{equation} \label{eq:11}
\langle \langle d_{\sigma}| d_{\sigma}^{\dagger}  \rangle \rangle_{\omega} = \sum_{\alpha \beta} \sum_{\gamma \delta} f_{\alpha \beta}^{\sigma} f_{\gamma \delta}^{\sigma \ast} \langle \langle A_{\alpha \beta}^{\sigma} | A_{\gamma \delta}^{\sigma \dagger}   \rangle \rangle_{\omega}.
\end{equation}
The EOM of a fermionic type double time GF $\langle \langle A | B \rangle \rangle_{\omega} $ reads
\begin{equation} \label{eq:12}
\omega \langle \langle A|B \rangle \rangle_{\omega} = \langle \{ A, B \} \rangle + \langle \langle [A, H]  | B \rangle \rangle_{\omega}.
\end{equation}
The average $\langle \{ A, B \} \rangle$ can be determined self-consistently from the GF's through the fluctuation-dissipation theorem,
\begin{equation} \label{eq:13}
    \langle BA \rangle = -\frac{1}{\pi} \int_{-\infty}^{\infty} \frac{1}{e^{\beta \omega} + 1} \text{Im} \langle \langle A|B\rangle\rangle_{\omega + i\eta} d\omega.
\end{equation}
Here $\eta = 0+$ is an infinitesimal positive number.

In the following, we apply the EOM to the SBO GF $ \langle \langle A_{\alpha \beta}^{\sigma} | A_{\gamma \delta}^{\sigma \dagger} \rangle \rangle_{\omega}$. Appropriate truncation approximations will be introduced at certain levels to truncate the hierarchical EOM and produce a closed self-consistent equations.
The first order EOM reads
\begin{equation} \label{eq:14}
   \omega  \langle\langle A_{\alpha \beta}^{\sigma}| A_{\gamma \delta}^{\sigma \dagger} \rangle \rangle_{\omega} = \langle \{ A_{\alpha \beta}^{\sigma},  A_{\gamma \delta}^{\sigma \dagger} \}  \rangle + \langle \langle \left[A_{\alpha \beta}^{\sigma}, H_{Aim} \right] | A_{\gamma \delta}^{\sigma \dagger} \rangle \rangle.
\end{equation}
To obtain the GF's, we use the commutation relation $\left[ A_{\alpha \beta}, H_0 \right] = \left( E_{\beta} - E_{\alpha} \right) A_{\alpha \beta}$ and the anti-commutation relation $\{ A_{\alpha \beta}, A_{\delta \gamma} \} = \delta_{\beta \delta} A_{\alpha \gamma} + \delta_{\alpha \gamma} A_{\delta \beta}$. Also, for fermionic type operators in different spaces, the following anti-commutation relations hold, $\{ A_{\alpha \beta}^{\sigma}, c_{k \sigma^{\prime} } \} = 0 $ and $\{ A_{\alpha \beta}^{\sigma}, c_{k \sigma^{\prime} }^{\dagger} \} = 0 $.
Using these relations, we get
\begin{eqnarray} \label{eq:15}
 && \left(\omega + E_{\alpha} - E_{\beta} \right) \langle \langle A_{\alpha \beta}^{\sigma}| A_{\gamma \delta}^{\sigma \dagger} \rangle \rangle_{\omega}
 = \delta_{\beta \delta} \langle A_{\alpha \gamma}\rangle + \delta_{\alpha \gamma} \langle A_{\delta \beta}\rangle
                                                        \nonumber \\
  && + \sum_{k \ge n+1, \sigma^{\prime}} V_{k}  \left[  \langle \langle  B_{\alpha \beta}^{\sigma \sigma^{\prime}} c_{k \sigma^{\prime}} |  A_{\gamma \delta} ^{\sigma \dagger}  \rangle \rangle_{\omega}
  - \langle \langle  D_{\alpha \beta}^{\sigma \sigma^{\prime}}c_{k \sigma^{\prime}}^{\dagger} | A_{\gamma \delta} ^{\sigma \dagger}  \rangle \rangle_{\omega} \right].   \nonumber \\
  &&
\end{eqnarray}
Here, $\langle A_{\alpha \gamma}\rangle$ and $\langle A_{\delta \beta} \rangle$ are thermal averages of SBO's which conserve electron number and spin. The two new operators $B_{\alpha \beta}^{\sigma \sigma^{\prime}}$   and $ D_{\alpha \beta}^{\sigma \sigma^{\prime}}$ are defined as
\begin{eqnarray} \label{eq:16}
&& B_{\alpha \beta}^{\sigma \sigma^{\prime} }  \equiv \{A_{\alpha \beta}^{\sigma} , d_{\sigma^{\prime}}^{\dagger} \}  = \sum_{\mu} f_{\mu \beta}^{\sigma^{\prime} \ast} A_{\alpha \mu} + \sum_{\nu} f_{\alpha \nu}^{\sigma^{\prime} \ast} A_{\nu \beta}
                                            \nonumber \\
&& D_{\alpha \beta}^{\sigma \sigma^{\prime}} \equiv \{ A_{\alpha \beta}^{\sigma} , d_{\sigma^{\prime}} \} = \sum_{\mu} f_{\mu \alpha}^{\sigma^{\prime} } A_{\mu \beta} + \sum_{\nu} f_{\beta \nu}^{\sigma^{\prime} } A_{\alpha \nu} .
                                               \nonumber \\
 &&
\end{eqnarray}
It is noted that $B_{\alpha \beta}^{\sigma \sigma^{\prime}}$  conserves $\hat{N}$, while $D_{\alpha \beta}^{\sigma \sigma^{\prime}} $ annihilates two electrons. Both are Grassmann even operators.

On the right hand side of Eq.(\ref{eq:15}) appear two higher order GF's. $ \langle \langle B_{\alpha \beta}^{\sigma \sigma^{\prime}} c_{k \sigma^{\prime}} |  A_{\gamma \delta} ^{\sigma \dagger}  \rangle \rangle_{\omega}$ is the propagator of an electron from the small system $H_0$ into the infinite residual bath, and $\langle \langle  D_{\alpha \beta}^{\sigma \sigma^{\prime}}  c_{k \sigma^{\prime}}^{\dagger} | A_{\gamma \delta} ^{\sigma \dagger}  \rangle \rangle_{\omega}$ describes electron propagating inside the small system but being accompanied by fluctuations of the residual bath. Both can be represented by GF's of the form $\langle \langle A_{\mu \nu} X_{k \sigma^{\prime}}| A_{\gamma \delta} ^{\sigma \dagger}  \rangle \rangle_{\omega}$, with $X= c$ or $X=c^{\dagger}$.  If one calculate EOM for these new GF's, even higher order GF's of the form $\langle \langle A_{\mu \nu} X_{k \sigma^{\prime}} X_{p \sigma^{\prime\prime}}| A_{\gamma \delta} ^{\sigma \dagger} \rangle \rangle_{\omega}$ will be produced.

Different truncation schemes have been proposed for the Anderson impurity models under GF EOM formalism. AAA is made for GF's involving one bath index.\cite{Hubbard1} The famous Lacroix approximation truncates the GF's involving three bath index and gives qualitative description for the Kondo resonance.\cite{Lacroix1} Luo {\it et al.} proposed the truncation scheme based on the concept of connected GF.\cite{Luo1} In their work the truncation at three bath index level produces improved results over Lacroix.

In Eq.(\ref{eq:15}), if we neglect the influence of the residual bath, i.e., if we  remove the last two terms, the SBO GF's are solved as
\begin{equation} \label{eq:17}
\langle \langle A_{\alpha \beta}^{\sigma}| A_{\gamma \delta}^{\sigma \dagger} \rangle \rangle_{\omega}
 = \frac{\delta_{\beta \delta} \langle A_{\alpha \gamma}\rangle + \delta_{\alpha \gamma} \langle A_{\delta \beta}\rangle  } {  \omega + E_{\alpha} - E_{\beta} }.
\end{equation}
Employing the fluctuation-dissipation theorem Eq.(\ref{eq:13}) and the SBO algebra Eq.(\ref{eq:8}), one gets the self-consistent equations for the averages, $\langle A_{\alpha \beta} \rangle  = \delta_{\alpha \beta} e^{-\beta E_{\alpha}}/Z_{0}$. $Z_0 = \sum_{\mu} e^{-\beta E_{\mu}}$ is the partition function of $H_0$. Putting the averages into the expression of SBO GF's and using Eq.(\ref{eq:11}), one gets
\begin{equation} \label{eq:18}
\langle \langle d_{\sigma}| d_{\sigma}^{\dagger} \rangle \rangle_{\omega}
 = \frac{1}{Z_0} \sum_{\mu \nu} \frac{ \left( e^{-\beta E_{\mu}} + e^{-\beta E_{\nu}}\right) f_{\mu\nu}^{\sigma} f_{\mu \nu}^{\sigma \ast} } { \omega + E_{\mu} - E_{\nu} }.
\end{equation}
It is the Lehmann representation of the impurity GF of $H_0$, as expected.

To take into account the influence of residual bath and improve over ED results, one needs to consider the EOM of the last two terms in Eq.(\ref{eq:15}). The EOM for $ \langle \langle B_{\alpha \beta}^{\sigma \sigma^{\prime}} c_{k \sigma^{\prime}} |  A_{\gamma \delta} ^{\sigma \dagger}  \rangle \rangle_{\omega}$
, taking a symmetric form to keep the possible particle-hole symmetry, is obtained as,
\begin{eqnarray} \label{eq:19}
  && \omega \langle \langle B_{\alpha \beta}^{\sigma \sigma^{\prime}} c_{k \sigma^{\prime}}| A_{\gamma \delta} ^{\sigma \dagger}  \rangle \rangle_{\omega}
                                           \nonumber \\
 &=& \frac{\omega}{2}\langle \langle \{ B_{\alpha \beta}^{\sigma \sigma^{\prime}}, c_{k \sigma^{\prime}} \}| A_{\gamma \delta} ^{\sigma \dagger}  \rangle \rangle_{\omega}         \nonumber \\
  &=& \langle \{ A_{\gamma\delta}^{\sigma \dagger},  B_{\alpha \beta}^{\sigma \sigma^{\prime}} c_{k \sigma^{\prime}} \} \rangle
  + \langle \langle \left[B_{\alpha \beta}^{\sigma \sigma^{\prime}}, H_0 \right] c_{k \sigma^{\prime}}| A_{\gamma \delta}^{\sigma \dagger}\rangle \rangle_{\omega}
                                             \nonumber \\
  &+&   \frac{1}{2} V_k \langle \langle \{ B_{\alpha \beta}^{\sigma \sigma^{\prime}}, d_{\sigma^{\prime}} \} |A_{\gamma \delta} ^{\sigma \dagger} \rangle \rangle_{\omega}
  + \epsilon_k \langle \langle B_{\alpha \beta}^{\sigma \sigma^{\prime}} c_{k \sigma^{\prime}} | A_{\gamma \delta} ^{\sigma \dagger}\rangle \rangle_{\omega}
                                                   \nonumber \\
  &+& \frac{1}{2} \sum_{p \sigma^{\prime \prime}} V_p \langle \langle \left[ B_{\alpha \beta}^{\sigma \sigma^{\prime}}, d_{\sigma^{\prime \prime}}^{\dagger}\right] \left[c_{p \sigma^{\prime \prime}}, c_{k \sigma^{\prime}} \right] | A_{\gamma \delta} ^{\sigma \dagger} \rangle \rangle_{\omega}
                                                 \nonumber \\
&+& \frac{1}{2} \sum_{p \sigma^{\prime \prime}} V_p \langle \langle \left[ B_{\alpha \beta}^{\sigma \sigma^{\prime}}, d_{\sigma^{\prime \prime}} \right] \left[ c_{k \sigma^{\prime}}, c_{p\sigma^{\prime \prime}}^{\dagger} \right]| A_{\gamma \delta} ^{\sigma \dagger} \rangle\rangle_{\omega}.
\end{eqnarray}

Since  the most important effect of the impurity-bath interaction is supposed to be contained in $H_0$, we may devise simple truncation approximations for the residual bath. From the practical requirement of a fast impurity solver, we also need to avoid the complications in the EOM part. In the following, we introduce AAA approximation which simplifies the EOM for spin-up GF by ignoring the commutator between the spin-down operators and the hybridization operators in $H$. That is, when studying the propagation of spin-$\sigma$ electrons, the fluctuations in spin-$\bar{\sigma}$ electron are frozen and treated as non-dynamical quantity.

The last two terms in Eq.(\ref{eq:19}) come from $\langle \langle \left[B_{\alpha \beta}^{\sigma \sigma^{\prime}}, H_{1} \right]c_{k \sigma^{\prime}}| A_{\gamma \delta} ^{\sigma \dagger}  \rangle \rangle_{\omega}$ and $\langle \langle c_{k \sigma^{\prime}}\left[B_{\alpha \beta}^{\sigma \sigma^{\prime}}, H_{1} \right]| A_{\gamma \delta} ^{\sigma \dagger}  \rangle \rangle_{\omega}$, which describe the fluctuation effect from the impurity- residue bath hybridization when electron is propagating. In the spirit of AAA, they are ignored. The second term on the right hand side of Eq.(\ref{eq:19}) comes from the dynamics of the small system $H_0$ accompanying the propagation of electrons. The EOM of this GF can lead to a chain of GF's of the type $\langle \langle O_{i} c_{k \sigma^{\prime}}| A_{\gamma \delta}^{\sigma \dagger}\rangle \rangle_{\omega} $ ($i=1,2,...$). This chain will get closed when the electron and spin conserving operator $O_{i}$ traverse the whole space of $A_{\alpha \beta}$. The more advanced solution taking these into account will be studied later. Here, we make the decoupling approximation based on the argument that $\Gamma_{2}(i\omega_n) $ is supposed to be small and can be treated as a perturbation,
\begin{eqnarray} \label{eq:20}
 && \langle \langle \left[B_{\alpha \beta}^{\sigma \sigma^{\prime}}, H_0 \right]c_{k \sigma^{\prime}}| A_{\gamma \delta}^{\sigma \dagger}\rangle \rangle_{\omega}
                                          \nonumber \\
 &&\approx  \langle \left[B_{\alpha \beta}^{\sigma \sigma^{\prime}}, H_0 \right] \rangle \langle \langle c_{k \sigma^{\prime}}| A_{\gamma \delta}^{\sigma \dagger}\rangle \rangle_{\omega}
                                           \nonumber \\
 && \approx  \langle \left[B_{\alpha \beta}^{\sigma \sigma^{\prime}}, H_0 \right] \rangle_{H_{0}} \langle \langle c_{k \sigma^{\prime}}| A_{\gamma \delta}^{\sigma \dagger}\rangle \rangle_{\omega}
                                   \nonumber \\
 && = 0.
\end{eqnarray}

For the first term in Eq.(\ref{eq:19}) $\langle \left[
A_{\gamma\delta}^{\sigma \dagger},  B_{\alpha \beta}^{\sigma
\sigma^{\prime}} \right] c_{k \sigma^{\prime}} \rangle$, it could be
calculated self-consistently from the GF of the type
$\langle \langle c_{k \sigma^{\prime}} | A_{\gamma\delta}^{\sigma
\dagger}\rangle \rangle_{\omega}$, which is connected to $\langle
\langle A_{\alpha\beta}^{\sigma} | A_{\gamma\delta}^{\sigma
\dagger}\rangle \rangle_{\omega}$ using EOM. For the moment, to make
the calculation simpler, we consider the following decoupling
approximation to this term, $ \langle A_{\gamma\delta}^{\sigma
\dagger} B_{\alpha \beta}^{\sigma \sigma^{\prime}} c_{k
\sigma^{\prime}} \rangle  \approx \langle A_{\gamma\delta}^{\sigma
\dagger}c_{k \sigma^{\prime}}  \rangle \langle
B_{\alpha\beta}^{\sigma \sigma^{\prime}}\rangle$ and $ \langle
B_{\alpha \beta}^{\sigma \sigma^{\prime}} c_{k \sigma^{\prime}}
A_{\gamma\delta}^{\sigma \dagger} \rangle  \approx  -\langle
B_{\alpha\beta}^{\sigma \sigma^{\prime}}\rangle \langle
A_{\gamma\delta}^{\sigma \dagger}c_{k \sigma^{\prime}}  \rangle$.
This leads to
\begin{equation} \label{eq:21}
    \langle \{ A_{\gamma\delta}^{\sigma \dagger},  B_{\alpha \beta}^{\sigma \sigma^{\prime}} c_{k \sigma^{\prime}} \} \rangle \approx 0.
\end{equation}

Finally, we ignore the spin flip process in Eq.(\ref{eq:19}) since they are much less important than the spin-conserving propogation, $\langle \langle B_{\alpha \beta}^{\sigma \sigma^{\prime}} c_{k \sigma^{\prime}}| A_{\gamma \delta} ^{\sigma \dagger}  \rangle \rangle_{\omega}  \approx \delta_{\sigma \sigma^{\prime}} \langle\langle B_{\alpha \beta}^{\sigma \sigma} c_{k \sigma}| A_{\gamma \delta} ^{\sigma \dagger}  \rangle \rangle_{\omega}$. With all these approximations, the second order EOM is simplified into
\begin{equation} \label{eq:22}
    \langle \langle B_{\alpha \beta}^{\sigma \sigma^{\prime}} c_{k \sigma^{\prime}}| A_{\gamma \delta} ^{\sigma \dagger}  \rangle \rangle_{\omega}
  \approx \delta_{\sigma \sigma^{\prime}} \frac{V_k}{2(\omega - \epsilon_k)} \langle \langle \{ B_{\alpha \beta}^{\sigma}, d_{\sigma} \} |A_{\gamma \delta} ^{\sigma \dagger} \rangle \rangle_{\omega},
\end{equation}
with $B_{\alpha\beta}^{\sigma} = B_{\alpha\beta}^{\sigma \sigma}$.
Carrying out similar EOM calculation for the other GF in Eq.(\ref{eq:15}), $\langle \langle  D_{\alpha \beta}^{\sigma \sigma^{\prime}}c_{k \sigma^{\prime}}^{\dagger} | A_{\gamma \delta} ^{\sigma \dagger}  \rangle \rangle_{\omega}$, and making approximations alike, we obtain
\begin{equation} \label{eq:23}
    \langle \langle c_{k \sigma^{\prime}}^{\dagger} D_{\alpha \beta}^{\sigma \sigma^{\prime}} | A_{\gamma \delta} ^{\sigma \dagger}  \rangle \rangle_{\omega}
  \approx \delta_{\sigma \sigma^{\prime}} \frac{-V_k}{2(\omega + \epsilon_k)} \langle \langle \{ D_{\alpha \beta}^{\sigma}, d_{\sigma}^{\dagger} \} |A_{\gamma \delta} ^{\sigma \dagger} \rangle \rangle_{\omega},
\end{equation}
with $D_{\alpha \beta}^{\sigma} = D_{\alpha \beta}^{\sigma \sigma}$

Putting Eq.(\ref{eq:22}) and (\ref{eq:23}) into Eq.(\ref{eq:15}), we obtain the closed set of equations for $\langle \langle A_{\alpha \beta}^{\sigma}| A_{\gamma \delta}^{\sigma \dagger} \rangle \rangle_{\omega}$ as
\begin{eqnarray} \label{eq:24}
 && \left(\omega + E_{\alpha} - E_{\beta} \right) \langle \langle A_{\alpha \beta}^{\sigma}| A_{\gamma \delta}^{\sigma \dagger} \rangle \rangle_{\omega}=\delta_{\beta \delta} \langle A_{\alpha \gamma}\rangle
                           \nonumber \\
&& + \delta_{\alpha \gamma} \langle A_{\delta \beta}\rangle
    + \frac{\Gamma_2(\omega)}{2} \sum_{\mu \nu} M^{\sigma}_{\alpha \beta, \mu\nu  }  \langle \langle  A_{\mu \nu}^{\sigma}|  A_{\gamma \delta} ^{\sigma \dagger}  \rangle \rangle_{\omega}.
                                \nonumber \\
  && -\frac{\Gamma_{2}(-\omega)}{2} \sum_{\mu \nu} N^{\sigma}_{\alpha \beta, \mu\nu} \langle \langle  A_{\mu \nu}^{\sigma}|  A_{\gamma \delta} ^{\sigma \dagger}  \rangle \rangle_{\omega}.
\end{eqnarray}
Here, $M^{\sigma}_{\alpha \beta, \mu\nu}$
and $N^{\sigma}_{\alpha \beta, \mu\nu}$ are defined by $\{ B_{\alpha, \beta}^{\sigma}, d_{\sigma} \} = \sum_{\mu, \nu} M^{\sigma}_{\alpha \beta, \mu\nu} A^{\sigma}_{\mu\nu}$ and $\{ D_{\alpha, \beta}^{\sigma}, d_{\sigma}^{\dagger} \} = \sum_{\mu, \nu} N^{\sigma}_{\alpha \beta, \mu\nu} A^{\sigma}_{\mu\nu}$, respectively. Their expressions are
\begin{eqnarray} \label{eq:25}
 M^{\sigma}_{\alpha \beta, \mu\nu} &=& \delta_{\alpha\mu}\left(\sum_{\tau} f^{\sigma \ast}_{\tau \beta} f^{\sigma}_{\tau\nu} \right) + \delta_{\beta \nu} \left( \sum_{\tau} f^{\sigma \ast}_{\alpha \tau} f^{\sigma}_{\mu \tau} \right)  \nonumber \\
   && + f^{\sigma \ast}_{\nu \beta} f^{\sigma}_{\mu\alpha} + f^{\sigma \ast}_{\alpha \mu} f^{\sigma}_{\beta\nu}  \nonumber \\
 N^{\sigma}_{\alpha \beta, \mu\nu} &=& \delta_{\alpha\mu}\left(\sum_{\tau} f^{\sigma \ast}_{\nu \tau} f^{\sigma}_{\beta \tau} \right) + \delta_{\beta \nu} \left( \sum_{\tau} f^{\sigma \ast}_{\tau \mu} f^{\sigma}_{\tau \alpha} \right)  \nonumber \\
   && + f^{\sigma \ast}_{\nu \beta} f^{\sigma}_{\mu\alpha} + f^{\sigma \ast}_{\alpha \mu} f^{\sigma}_{\beta\nu}.
\end{eqnarray}
The coefficients $M_{\alpha\beta, \mu\nu}^{\sigma}$ and $N_{\alpha\beta, \mu\nu}^{\sigma}$ satisfy two sets of exact relations which can be used for numerical test. For details see Appendix C.

Multiplying $f_{\gamma \delta}^{\sigma \ast}$ to Eq.(\ref{eq:24}) and summing over $\gamma$ and $\delta$, we get
\begin{eqnarray} \label{eq:26}
&& \left(\omega + E_{\alpha} - E_{\beta} \right) \langle \langle A_{\alpha \beta}^{\sigma}| d_{\sigma}^{\dagger} \rangle \rangle_{\omega} = \sum_{\gamma} f_{\gamma \beta}^{\sigma \ast} \langle A_{\alpha \gamma}\rangle        \nonumber \\
  && + \sum_{\delta} f_{\alpha \delta}^{\sigma \ast} \langle A_{\delta \beta}\rangle
 + \frac{\Gamma_2(\omega)}{2} \sum_{\mu \nu} M_{\alpha \beta, \mu\nu
  }  \langle \langle  A_{\mu \nu}^{\sigma}|  d_{\sigma} ^{\dagger}  \rangle \rangle_{\omega}      \nonumber \\
  &&- \frac{\Gamma_{2}(-\omega)}{2} \sum_{\mu \nu} N_{\alpha \beta, \mu\nu }  \langle \langle  A_{\mu \nu}^{\sigma}|  d_{\sigma} ^{\dagger}  \rangle \rangle_{\omega}.
\end{eqnarray}
Eq.(\ref{eq:26}) is a set of linear equations about $D^2$ unknowns $\{ \langle \langle A_{\alpha \beta}^{\sigma}| d_{\sigma}^{\dagger} \rangle \rangle_{\omega} \}$. For each $\omega$, it can be solved using iterative methods. Parallel computation can be used for different $\omega$ to reduce the computation time.
It is the core equation on which our following calculations and discussions are based.

There are $D^2$ unknown averages $ \{\langle A_{\alpha \beta} \rangle \}$ in the inhomogeneous term of Eq.(\ref{eq:26}). They need to be solved self-consistently using GF's via the fluctuation-dissipation theorem Eq.(\ref{eq:13}). An equation about the averages can be obtained from each GF's shown above,
\begin{eqnarray} \label{eq:27}
    &&  \langle  d_{\sigma}^{\dagger}  A_{\alpha \beta}^{\sigma}\rangle = \sum_{\delta} f_{\alpha \delta}^{\sigma \ast} \langle A_{\delta \beta}\rangle
                               \nonumber \\
&& =  -\frac{1}{\pi} \int_{-\infty}^{\infty} \frac{1}{e^{\beta \omega} + 1} \text{Im} \langle \langle  A_{\alpha \beta}^{\sigma} |d_{\sigma}^{\dagger} \rangle\rangle_{\omega + i\eta}  d\omega.
\end{eqnarray}
Thus we obtain $D^2$ equations for the same number of unknown averages, which can be solved iteratively.
In principle, from Eq.(\ref{eq:24}) one could construct more than sufficient self-consistent equations for the averages $ \{\langle A_{\alpha \beta} \rangle \}$. However, it is noted that neither linear independence nor consistency is guaranteed in these linear equations, due to the approximations made for GF. There have been proposals for systematic ways of constructing sufficient and consistent equations,\cite{Haley2} but in general this is still an open problem in the GF EOM approach.

In order to keep our calculation scheme as simple as possible, we adopt the simplest approximation for calculating the averages,
\begin{equation}  \label{eq:28}
\langle A_{\alpha \beta}\rangle \approx \langle A_{\alpha \beta}\rangle^{0} = \delta_{\alpha \beta} e^{-\beta E_{\alpha}}/Z_0
\end{equation}
The advantage of using $H_0$ result for $\langle A_{\alpha \beta}\rangle$ is that no self-consistent calculation is required and the sum rule is fulfilled exactly. The shortcoming is that the finite size effect of $H_0$ will enter the final results and discontinuous change of averages and GF's will occur when parameters such as $\mu$ or $U$ are tuned continuously at $T=0$. For our calculation below, we stay at half-filling and this shortcoming has no effect.

Putting the approximate averages into Eq.(\ref{eq:26}), we get the linear equations for the final GF's as
\begin{eqnarray} \label{eq:29}
  && \sum_{\mu\nu} \mathcal{K}_{\alpha\beta, \mu\nu}(\omega) \langle \langle  A_{\mu \nu}^{\sigma}| d_{\sigma}^{\dagger} \rangle\rangle_{\omega}
     = f_{\alpha\beta}^{\sigma \ast} \left[ \langle A_{\alpha \alpha} \rangle^{0} +\langle A_{\beta \beta}\rangle^{0} \right],   \nonumber \\
  &&
\end{eqnarray}
where
\begin{eqnarray}  \label{eq:30}
  && \mathcal{K}_{\alpha\beta, \mu\nu} (\omega)=  \nonumber \\
  && \left(\omega + E_{\alpha} - E_{\beta} \right)\delta_{\alpha \mu} \delta_{\beta \nu}  - \frac{1}{2}\Gamma_{2}(\omega) M_{\alpha\beta, \mu\nu} + \frac{1}{2}\Gamma_{2}(-\omega) N_{\alpha\beta, \mu\nu}.   \nonumber \\
  &&
\end{eqnarray}
Finally, the local GF of the impurity site is obtained as
\begin{equation} \label{eq:31}
   \langle \langle d_{\sigma} | d_{\sigma}^{\dagger} \rangle \rangle_{\omega} = \sum_{\alpha \beta} f_{\alpha \beta}^{\sigma} \langle \langle A_{\alpha \beta}^{\sigma} | d_{\sigma}^{\dagger} \rangle \rangle_{\omega}.
\end{equation}

Before we present the numerical results, some discussions about the limiting cases are in order. First, in the limit $n_{s}=0$, equation Eq.(\ref{eq:29}) is equivalent to AAA, as being proved in Appendix A. This is because at $n_{s}=0$, $\Gamma_{2}(\omega) = \Gamma(\omega)$ and the full impurity-bath hybridization is treated at AAA level. In the other limit $n_{s}=\infty$, the full continuous bath is contained in $H_0$ and the residual hybridization $\Gamma_{2}(\omega)=0$. As shown in Eq.(\ref{eq:18}), the exact GF obtains. We therefore expect a smooth interpolation between AAA and ED as $n_{s}$ increases from zero. Especially, the coherent quasi-particle features missing in the AAA spectral function should be produced for $n_{s} \geq 1$.

Second, the set of equations Eq.(\ref{eq:26}) becomes exact in the limit $U=0$, as proved in Appendix B. In the other limit of weak hybridization $\Gamma(\omega)=0$, our result is exact because the local impurity part is contained in $H_0$.

\section{Results}
\subsection{Anderson impurity model}
We first present the results of our impurity solver for the Anderson impurity model Eq.(2). For this model, we use a Lorentzian hybridization function 
\begin{equation}  \label{eq:32}
    \Delta(\omega) = \frac{\Delta \omega_{c}^2}{\omega^2 + \omega_c^2}.
\end{equation}
$\omega_c = 1$ sets the unit of energy and $\Delta$ describes the strength of the hybridization. The corresponding Matsubara hybridization function reads
\begin{equation}  \label{eq:33}
    \Gamma(i \omega_n) = \frac{\pi \Delta \omega_{c}}{i\omega_n + i\omega_c Sgn (\omega_n) }.
\end{equation}

In order to parametrize the discrete bath degrees of freedom in $H_0$, we define the distance $d$ between the input hybridization function $\Gamma(i\omega_n)$ and that of the small system $\Gamma_{1}(i\omega_n)$,
\begin{equation}   \label{eq:34}
   d = \frac{1}{N} \sum_{n} \left[ | \Gamma(i\omega_n) - \Gamma_{1}(i\omega_n)|^{2}/ \omega_n^{s} \right],
\end{equation}
with $N$ the number of summed Matsubara frequencies.
$d$ is then minimized with respect to variables $\epsilon_{k}$ and $V_{k}$ ($k=1,2,...,n_s$) use conjugate gradient method. For the number of bath sites $n_{s} \leq 2$, our results are sensitive to $s$ value. We checked that $s \geq 3$ always leads to too small $\epsilon_{k}$'s and below we use $s=2$.\cite{Hafermann1}

Fig.1 depicts the local density of states at $T=0.2\pi\Delta$ and for different $U$'s, obtained using $n_{s}=1$ at the particle-hole symmetric point $\mu=\frac{U}{2}$. At $U = 0$, the density of states coincides with the exact one, with the height of the peak $\rho(0)=1/(\Delta\pi^2)$. As $U$ increases, the weight of the central peak transfers to higher energies, leading to a sharp central Kondo peak. The two broad peaks around $U=\pm U/2$ are the incoherent Hubbard peaks, forming the well known three-peak structure. All the spectral function curves are non-negative and the sum rule $\int \rho(\omega) d\omega = 1$ is fulfilled. This shows that our approach has no causality problem as in the strong coupling expansion.\cite{Dai1} 

As $U$ increases, $\rho(0)$ stay very close to the $U=0$ value. This is in qualitative agreement with the Fermi liquid behavior. 
At the quantitative level, however, the results in Fig.1 deviate from NRG results (not shown here) in that $\rho(0)$ decrease too slowly with increasing $U$ at $T=0.2\pi\Delta$. This is because the Kondo resonance obtained here is mainly from the hybridization between the impurity and the single bath level contained in $H_0$. The whole low energy bath is only represented by this bath level at $\epsilon=0$. The approximate way of calculating $\langle A_{\alpha\alpha} \rangle$ in Eq.(\ref{eq:28}) also makes our result less accurate in the temperature dependence.

\begin{figure}[t!]
\vspace{-0.0in}
\begin{center}
\includegraphics[width=4.9in, height=3.7in, angle=0]{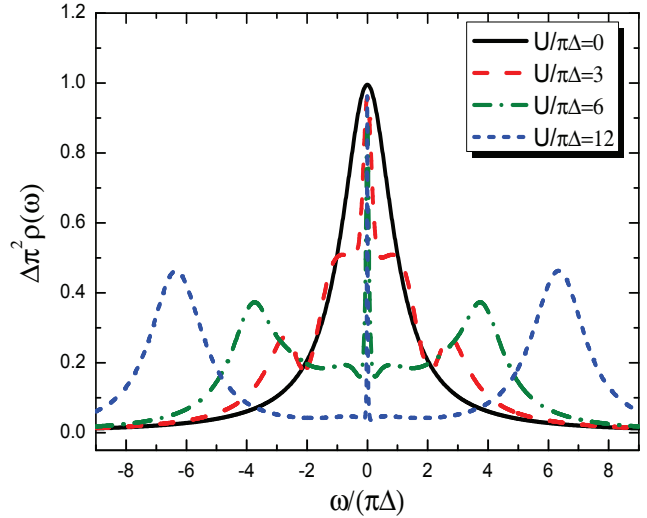}
\vspace*{-1.5cm}
\end{center}
\caption{The spectral function of Anderson impurity model at different $U$'s and temperature $T=0.2 \pi \Delta$. They are obtained using $n_{s}=1$ for Lorentzian hybridization function with $\pi \Delta=0.02$. The broadening parameter is $\eta = 10^{-4}$.}
\end{figure}

The quasi-particle weight of the impurity electron is defined as 
\begin{equation}   \label{eq:35}
z = \left[1 - \frac{ \partial {\text Re}\Sigma_{\sigma}(\omega) } { \partial \omega } \right]^{-1} \approx \left[1 - \frac{ {\text Im}\Sigma_{\sigma}(i \omega_0) } { \omega_0 } \right]^{-1}.
\end{equation}
$\Sigma_{\sigma}(i\omega_n)$ is the Matsubara self-energy
\begin{equation}   \label{eq:36}
\Sigma_{\sigma}(i\omega_n) = \left[ \langle \langle d_{\sigma} | d_{\sigma}^{\dagger} \rangle \rangle^{0}_{i\omega_n} \right]^{-1} - \left[ \langle \langle d_{\sigma} | d_{\sigma}^{\dagger} \rangle \rangle_{i\omega_n} \right]^{-1},
\end{equation}
where $\langle \langle d_{\sigma} | d_{\sigma}^{\dagger} \rangle \rangle^{0}_{i\omega_n}$ is the impurity Matsubara Green's function at $U=0$. In Eq.(\ref{eq:35}), $z$ is obtained either by evaluating the partial differentiation at the smallest pole of GF (Ref.\onlinecite{Bulla2,Potthoff1}), or from Matsubara self-energy $\Sigma(i\omega_{0})$. Here we find that due to the finite energy resolution of small $n_s$, the latter approach gives better result, as shown in Fig.2. $z$ decreases monotonically from $1$ at $U=0$ to zero in large $U$ limit, in qualitative agreement with the NRG result. However, due to the same reason of small $n_{s}$, our $z$ value sensitively depends on $T$. Only at intermediate temperature $T=0.1\pi\Delta$ is our $z$ close the NRG data which is obtained at $T=10^{-5}$. At low $T$ our result is small than NRG result.

\begin{figure}[t!]
\vspace{-0.0in}
\begin{center}
\includegraphics[width=5.5in, height=4.0in, angle=0]{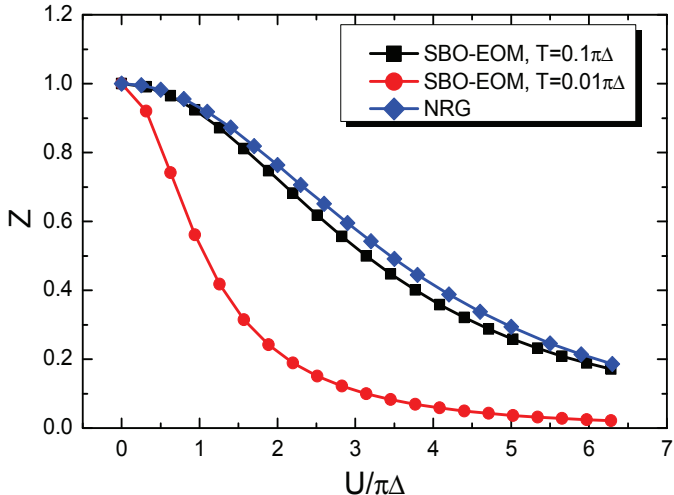}
\vspace*{-1.5cm}
\end{center}
\caption{Quasi-particle weight $z$ for the Anderson impurity model with Lorentzian hybridization function at $\pi\Delta=0.02$. It is obtained using $n_{s}=1$ at $T=0.1\pi\Delta$ (squares) and $T=0.01\pi\Delta$ (dots), respectively. The diamonds with guiding line is the NRG result at $T=10^{-5}$, obtained using $\Lambda=2$ and keeping $256$ states.}
\end{figure}

We compare the spectral function $\rho(\omega)$ and the Matsubara GF $G(i\omega_n)$ from different number of bath sites $n_s$ in $H_0$. The results are shown in Fig.3 at $U=3\pi\Delta$. In Fig.3(a), we see that for $n_{s}=0$, there is no Kondo resonance at all because the full bath is treated at AAA level. For $n_{s}=1$, a sharp resonance appears at $\omega=0$. For $n_{s}=2$, the sharp resonance is replaced with a broad peak. NRG result at the same temperature is shown for comparison. Qualitative agreement between NRG and SBO-EOM results are observed, but systematic convergence with increasing $n_s$ is not obvious up to $n_s=2$ here. In Fig.3(b), Matsubara GF is compared with ED results using $n_s=6$ which is convergent already. It is seen that the SBO-EOM result achieves quantitative agreement with ED at $n_{s}=1$ level already.

We also compared $\rho(\omega)$ with NRG data for larger $U$ values. 
There, $n_s=1$ gives qualitatively good result, but for $n_s=2$, instead of Kondo resonance at $\omega=0$, two sharp peaks appear around some $\omega_p \neq 0 $, which is actually related to the position of poles of $\Gamma_1(\omega)$, i.e., $\epsilon_k$'s in $H_0$ from fitting. The poles in $\Gamma_{1}(\omega)$ will appear in the final results, by entering $\mathcal{K}_{\alpha\beta, \mu\nu} (\omega)$ in Eq.(\ref{eq:30}) through $\Gamma_{2}(\omega)$. In this sense, the sharp peak at $\omega=0$ observed in Fig.1 should also be closely related to the fact that $\epsilon_k=0$ in $H_0$. Although the convergence of both $\rho(\omega)$ and $G(i\omega_n)$ in the large $n_{s}$ limit is guaranteed by the formula, qualitative correctness of $\rho(\omega)$ is limited to $n_s=1$ for the moment. For $n_s=1$, our calculation takes seconds on a PC. As $n_s$ increases, the computational cost increases exponentially and the efficiency advantage of the present solver will be lost very quickly. Therefore, we confine ourself to $n_{s} = 1$ when we use this solver for DMFT applications.

\begin{figure}[t!]
\vspace{-0.0in}
\begin{center}
\includegraphics[width=5.5in, height=4.3in, angle=0]{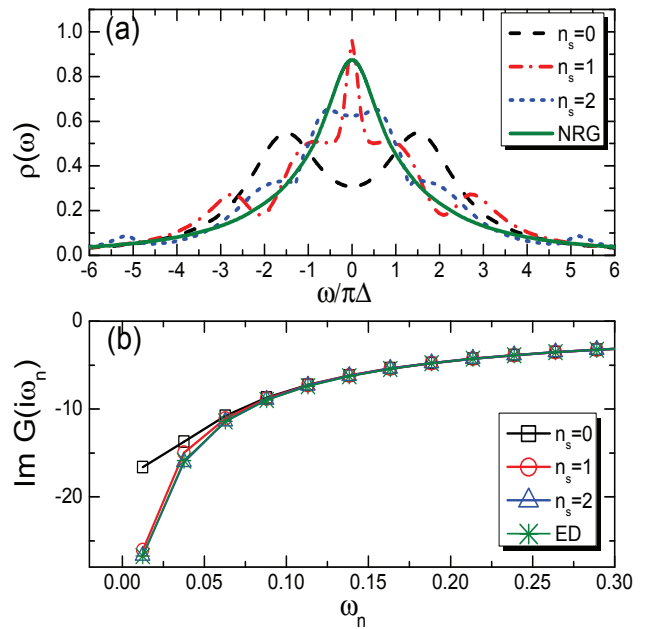}
\vspace*{-1.0cm}
\end{center}
\caption{The local density of states (a) and Matsubara Green's function (b) of Anderson impurity model, for bath numbers $n_s =0$ (square), $1$ (circle), and $2$ (triangle). $s=2$ is used in (b). Both are for Lorentzian hybridization at parameters $U=3 \pi\Delta$, $T=0.2 \pi\Delta$, and $\pi\Delta=0.02$. The solid line in (a) and the stars in (b) are NRG and ED results at same parameters, respectively. NRG result is obtained using full density matrix algorithm at $\Lambda=2$ and keeping $256$ states.}
\end{figure}

\subsection{Hubbard model}
  In this section, we combine the impurity solver with DMFT calculation to study the paramagnetic phase of the single band Hubbard model on the Bethe lattice. Due to the simplicity of the DMFT equation in this case, it is an ideal system for benchmarking our method.
The Hamiltonian of Hubbard model reads,
\begin{equation}     \label{eq:37}
   H = -\sum_{\langle i,j \rangle} t_{ij} c_{i\sigma}^{\dagger} c_{j\sigma} + U n_{\uparrow}n_{\downarrow} - \mu \sum_{\sigma} n_{\sigma}.
\end{equation}
For Bethe lattice with infinite coordination number, the bare density of states reads,
\begin{equation}     \label{eq:38}
   D(\omega) = \frac{2}{\pi W^2} \sqrt{W^2 - \omega^2}.
\end{equation}
Putting this into the standard DMFT self-consistent equations, it leads to the equation
\begin{equation}     \label{eq:39}
   \Gamma(\omega) = \frac{W^2}{4} G_{\sigma}(\omega).
\end{equation}
The half bandwidth $W$ is set as the energy unit $W=1$. 

  First we investigate the local density of states using $n_s=1$. Its evolution with $U$ is shown in Fig.4. At $U=0$, the exact $\rho(\omega)$ is obtained, as expected from Appendix B. When we increase U, the weight of the local density of states at small frequency transfers to higher energies, leading to a sharp middle peak, together with pronounced upper and lower Hubbard peaks at $\omega \sim \pm U/2$. This trend is consistent with previous studies using many well established methods, such as the iterative perturbation theory (IPT) and the quantum Monte Carlo (QMC) method.\cite{Georges1}  At the same time, $\rho(\omega=0)$ declines even for $U$ much smaller than the Mott transition point $U_c$. This is contradict to the Fermi liquid theory at such a low temperature. When $U$ is close to $1.2$, a gap opens with a small resonance peak standing inside the gap. The gap is fully formed at about $U_c \approx 1.3W$, completing the Mott metal-insulator transition. 
  
It is seen that our results qualitatively resemble the ones of IPT and QMC. However, there are important quantitative difference. The obtained critical $U_c \approx 1.3W$ in our calculation is much smaller than $U_{c}^{NRG} \approx 2.94W$ from DMFT(NRG)\cite{Bulla3} and $U_{c}^{TS} \approx 3.0W$ from the two-site DMFT\cite{Bulla2}. Compared to the AAA results $U_{c}^{AAA} = W$, our result is between DMFT(NRG) and AAA values. This shows that using $n_{s}=1$, our solver is an improvement over AAA but is not as accurate as the variational method that employs one bath site in an optimal way\cite{Bulla2}.

\begin{figure}[t!]
\vspace{-0.0in}
\begin{center}
\includegraphics[width=4.8in, height=3.4in, angle=0]{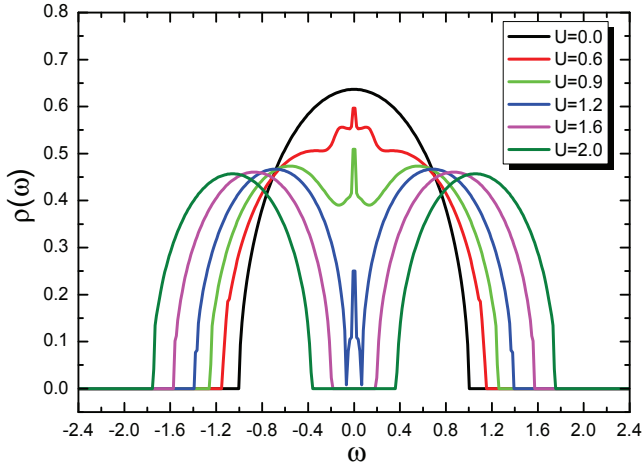}
\vspace*{-1.0cm}
\end{center}
\caption{The density of state of Hubbard model on Bethe lattice, obtained using SBO-EOM with $n_s=1$ as impurity solver. From top to bottom at $\omega=0$, $U=0.0, 0.6, 0.9, 1.2, 1.6$ and $2.0$, respectively. Here $\mu=U/2$ and $T=0.02$, $W=1.0$.}
\end{figure}

Fig.5 shows $\rho(0)$ as a function of $U$ at $T=0.02W$ which is much lower than the critical temperature. $\rho(0)$ decreases monotonically and drop to zero at $U_c \approx 1.3W$. The Mott transition in infinite spatial dimensions is of first order at finite temperatures. There is a finite regime $U_{c1} < U < U_{c2}$ where both metal and insulator coexist and there is metastable structure in thermodynamics.\cite{Tong1} In the inset of Fig.5, we show that indeed our impurity solver can describe such first order phase transition, although the obtained coexisting regime is much smaller than what was found in previous studies. As $U$ decreases across the transition point, $\rho(0)$ jumps from zero to a finite value at different $U_c$'s, giving out a coexistence regime of $U$.

\begin{figure}[t!]
\vspace{-0.0in}
\begin{center}
\includegraphics[width=4.5in, height=3.2in, angle=0]{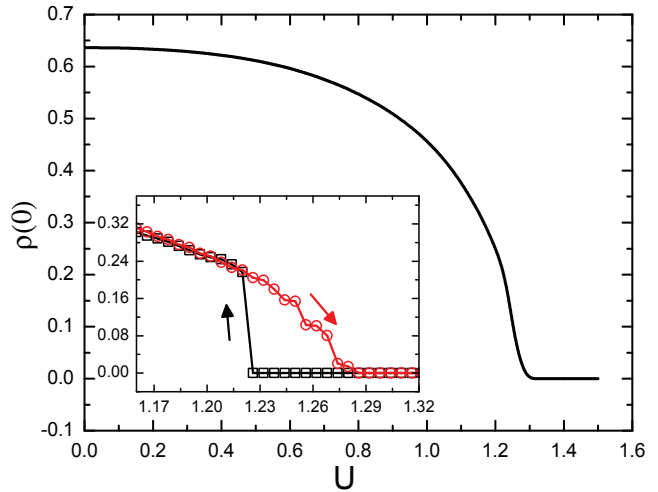}
\vspace*{-1.0cm}
\end{center}
\caption{$\rho(0)$ as the function of $U$ at $T=0.02$. Inset: the hysteresis near the Mott transition point.}
\end{figure}

\section{Discussions}

We first discuss the convergence of our results with respect to increasing $n_{s}$, the number of bath sites in $H_0$. Comparing $\rho(\omega)$ among $n_{s} = 0$ to $2$, it is found that apparently it does not show the expected convergence. This is mainly because the finite size effect in $H_0$ will influence the final result, by adding poles in the residual hybridization function $\Gamma_{2}(\omega)$. Therefore, for the real frequency spectral function, our solver produces an effective broadening for the ED results. On the other hand, the Matsubara GF converges quickly to the exact curve as $n_{s}$ increases. Note that by construction, our method will be exact in the large $n_{s}$ limit.
The way of convergence can be understood from Eq.(\ref{eq:30}): at those frequency points where the fitting is perfect $\Gamma_{1}(\omega_i)= \Gamma(\omega_i)$, our result equals to ED result because $\Gamma_2(\omega_i) = 0$. For a given $n_s$, it is empirically observed that there are $n_s$ different $\omega_i$'s where $\Gamma_{2}(\omega_i)=0$, meaning $n_s$ different frequencies where $G(\omega_i)$ is identical to ED values. With increasing $n_s$, $G(\omega)$ thus approaches the ED result consistently.

Second, the temperature effect is partly taken into account in out calculation (not shown here). We have observed that the sharp Kondo peak for the Anderson impurity model disappears as $T$ increases, leading to a broad featureless maximum. This is consistent with the expected behavior when $T$ goes above the Kondo temperature $T_k$. However, the hight of the spectral function $\rho(0)$ is higher than NRG result in this process. This problem, as well as the temperature dependence of $z(U)$ curve can be partly traced back to the approximations used to simplify the self-consistent calculation of the averages $\langle A_{\alpha \beta}\rangle$, Eq.(\ref{eq:28}). We expect that by replacing this approximation with the full self-consistent solution, the temperature dependence will be improved.

Third, we discuss possible improvement that we can make in the future.
In the approximations of GF, the spin flip process ignored in the
derivation can be taken into account without additional numerical effort. On the computation side, our calculations for $n_{s}=2$ takes only seconds on a PC. Increasing $n_s$ will lead to much slower computation and larger storage requirement. However, we expect that the odd number of $n_ｓ=3$ and $5$ can be reached with ease, which may give better description of both the Kondo resonance and the Hubbard peaks. The check of the convergence of results with respect to odd $n_{s}$ will be done later. Also, with full self-consistent calculation of the averages $\langle A_{\alpha \beta} \rangle$, we expect better description of the temperature dependence will be achieved.

\section{Summary}
In this work, we develop and benchmark a new impurity solver for DMFT, based on the EOM of double time GF of SBO's. Applying this method to the Anderson impurity model, we obtain qualitatively correct results for the real frequency spectral function $\rho(\omega)$, the quasi-particle weight $z$, and the Matsubara GF. We applied this method to the one band Hubbard model on Bethe lattice through DMFT and obtained qualitative description for the Mott transition. Directions of further improvement in the future is discussed.


\section{Acknowledgements}
This work is supported by the 973 Program of China (2012CB921704), National Natural Science Foundation of China
(11374362), Fundamental Research Funds for the Central 
Universities, and the Research Funds of 
Renmin University of China.

\appendix{}

\section{ Solution for $n_{s}=0$ case}
In this appendix, we prove that at $n_{s}=0$, Eq.(\ref{eq:29})-(\ref{eq:31}) is equivalent to AAA . For $n_{s}=0$, $H_0= Un_{\uparrow}n_{\downarrow} -\mu\sum_{\sigma} n_{\sigma}$. Hence $\Gamma_1(\omega)=0$ and $\Gamma_2(\omega)=\Gamma(\omega)$. The $4$ eigen states of $H_0$ are denoted as
\begin{eqnarray}   \label{eq:A1}
  && |1 \rangle = |\uparrow \rangle,   \,\,\,\,\, |2 \rangle = |\downarrow \rangle,   \nonumber \\
  && |3 \rangle = |0 \rangle,     \,\,\,\,\,\,   |4 \rangle = |\uparrow \downarrow \rangle.
\end{eqnarray}
The corresponding eigen energies are $E_{1}=E_{2}=-\mu$, $E_{3}=0$, and $E_{4}=U-2\mu$. We decompose $d_{\sigma}$ into SBO and obtain $d_{\uparrow} = A_{31}+A_{24}$ and $d_{\downarrow} = A_{32}-A_{14}$.
The nonzero coefficients $f^{\sigma}_{\alpha \beta}$ are
\begin{eqnarray}   \label{eq:A2}
  && f^{\uparrow}_{31} = f^{\uparrow}_{24}=1,  \nonumber \\
  && f^{\downarrow}_{14} = -1, \,\,\,\,\, f^{\downarrow}_{32}=1.  \nonumber \\
\end{eqnarray}
Putting these into Eq.(\ref{eq:25}), we get the only non-zero element among $M^{\uparrow}_{31,\mu\nu}$ as $M^{\uparrow}_{31,31}=2$, and that among $M^{\uparrow}_{24,\mu\nu}$ as $M^{\uparrow}_{24,24}=2$. Also one obtains $N^{\uparrow}_{31,\mu\nu} = N^{\uparrow}_{24,\mu\nu} = 0$ for all $(\mu\nu)$'s.

With these information, Eq.(\ref{eq:26}) becomes
\begin{equation}   \label{eq:A3}
  \langle \langle  A_{\alpha \beta}^{\sigma}| A_{\gamma \delta}^{\sigma \dagger} \rangle\rangle_{\omega}
   = \frac{ \delta_{\beta \delta}\langle A_{\alpha \gamma} \rangle +  \delta_{\alpha \gamma}\langle A_{\delta \beta} \rangle }{ \omega + E_{\alpha} -E_{\beta} - \Gamma(\omega) }.
\end{equation}
Using the fact that $A_{44}+A_{22}=d_{\downarrow}^{\dagger}d_{\downarrow}$ and $A_{11}+A_{33}=1 - d_{\downarrow}^{\dagger}d_{\downarrow}$, one obtains the final GF at $n_{s}=0$ as
\begin{equation}   \label{eq:A4}
  \langle \langle d_{\uparrow} | d_{\uparrow}^{\dagger} \rangle\rangle_{\omega}
   = \frac{ 1-\langle n_{\downarrow} \rangle}{\omega + \mu - \Gamma(\omega)} + \frac{\langle n_{\downarrow} \rangle}{ \omega + \mu - U -\Gamma(\omega) }.
\end{equation}
summarizing the spin-$\uparrow$ and -$\downarrow$ case, we obtain
\begin{equation}   \label{eq:A5}
  \langle \langle d_{\sigma} | d_{\sigma}^{\dagger} \rangle\rangle_{\omega}
   = \frac{ 1-\langle n_{\bar{\sigma}} \rangle}{\omega + \mu - \Gamma(\omega)} + \frac{\langle n_{\bar{\sigma}} \rangle}{ \omega + \mu - U -\Gamma(\omega) }.
\end{equation}
This is exactly the expression of AAA if the average $\langle n_{\bar{\sigma}}\rangle$ is calculated self-consistently. If we use its $H_0$ value $\langle n_{\bar{\sigma}}\rangle^{0}$ as was done in Eq.(\ref{eq:28}), it is not the full self-consistent AAA in general. But at exact half-filling $\langle n_{\bar{\sigma}}\rangle$ takes its universal value $1/2$, the resulting $G(\omega)$ recovers AAA again.

\section{Exact solution at $U=0$ limit}
In this appendix, we prove that Eq.(\ref{eq:26}) gives exact GF in the limit $U=0$.

In the case $U=0$, $H_0$ can be diagonalized on the single-particle level and we suppose that after diagonalization, $H_0$ reads

\begin{equation}   \label{eq:B1}
   H_0 = \sum_{s \sigma} \epsilon_{s} a_{s \sigma}^{\dagger} a_{s \sigma}.
\end{equation}
Here the operator $a_{s \sigma}$ is the annihilation operator of an electron in the molecular orbital $s$ and with spin $\sigma$.
The impurity annihilation operator $d_{\sigma}$ can be expanded as $d_{\sigma} = \sum_{s} \alpha_{s \sigma} a_{s\sigma}$, with $\sum_{s} |\alpha_{s \sigma}|^2 = 1$. The eigen energies $\{ E_{\mu} \}$'s are the sum of the occupied single particle levels. The SBO $A_{\alpha \beta}$ and the excitation energy $E_{\alpha}-E_{\beta}$ are defined similarly as in Eq.(\ref{eq:9}). In this case, considering that $\left[a_{s\sigma}, H_0 \right] = \epsilon_{s \sigma} a_{s \sigma}$, the expansion of $a_{s\sigma}$ in terms of SBO's $\{ A_{\alpha \beta}^{\sigma} \}$ reads
\begin{equation} \label{eq:B2}
   a_{s\sigma} =  \widetilde{\sum}_{\mu \nu} h_{\mu \nu}^{s \sigma} A_{\mu \nu}^{\sigma}.
\end{equation}
Here $ \widetilde{\sum}_{\mu \nu}$ denotes the summation with the constraint $E_{\nu} - E_{\mu} = \epsilon_{s}$, i.e., the  summation of degenerate excitations only.

Multiplying Eq.(\ref{eq:26}) with $h_{\alpha \beta}^{s \sigma}$ and summing over $\alpha$ and $\beta$ under the constraint $E_{\beta}-E_{\alpha} = \epsilon_{s}$, one obtains
\begin{eqnarray} \label{eq:B3}
 && \left(\omega - \epsilon_{s} \right) \langle \langle a_{s \sigma} | d_{\sigma}^{\dagger} \rangle \rangle_{\omega}
                           \nonumber \\
& = & \langle \{ a_{s \sigma}, d_{\sigma}^{\dagger}  \} \rangle
   + \frac{1}{2}\Gamma_2(\omega) \langle \langle  \{ \{ a_{s \sigma}, d_{\sigma}^{\dagger} \}, d_{\sigma} \} |  d_{\sigma} ^{\dagger}  \rangle \rangle_{\omega}
                           \nonumber \\
&& - \frac{1}{2} \Gamma_{2}(-\omega)  \langle \langle  \{ \{ a_{s \sigma}, d_{\sigma} \}, d_{\sigma}^{\dagger} \} |  d_{\sigma}^{\dagger}  \rangle \rangle_{\omega} .
\end{eqnarray}
Here we have used the relation $\sum_{\mu\nu} M_{\alpha\beta, \mu\nu}A_{\mu\nu}^{\sigma} = \{B_{\alpha \beta}^{\sigma}, d_{\sigma} \} = \{ \{A_{\alpha \beta}^{\sigma}, d_{\sigma}^{\dagger} \}, d_{\sigma} \} $, and $\sum_{\mu\nu} N_{\alpha\beta, \mu\nu}A_{\mu\nu}^{\sigma}= \{D_{\alpha \beta}^{\sigma}, d_{\sigma}^{\dagger} \} = \{ \{A_{\alpha \beta}^{\sigma}, d_{\sigma} \}, d_{\sigma}^{\dagger} \} $.

Using the definition of $a_{s \sigma}$, Eq.(\ref{eq:B3}) leads to
the equation for GF as
\begin{equation} \label{eq:B5}
   \langle \langle  d_{\sigma} |  d_{\sigma} ^{\dagger}  \rangle \rangle_{\omega} = \left( \sum_{s} \frac{|\alpha_{s\sigma}|^2}{\omega - \epsilon_s} \right) \left[1 +  \Gamma_{2}(\omega)    \langle \langle  d_{\sigma} |  d_{\sigma} ^{\dagger}  \rangle \rangle_{\omega} \right].
\end{equation}
Note that for $U=0$ case, the GF of the small system $H_0$ can be expressed as
\begin{equation}
 G_{0}(\omega) = \sum_{\alpha} \frac{|\alpha_{s\sigma}|^2}{\omega - \epsilon_s}   = \frac{1}{\omega + \mu - \Gamma_{1}(\omega)}.
\end{equation}
Putting this equation into Eq.(\ref{eq:B5}), we finally recover the exact impurity GF of $H_{Aim}$ at $U=0$,
\begin{equation} \label{eq:B6}
  \langle \langle d_{\sigma} | d_{\sigma}^{\dagger} \rangle \rangle_{\omega} = \frac{1}{G_{0}^{-1}(\omega) - \Gamma_2(\omega)} = \frac{1}{\omega + \mu - \Gamma(\omega) }.
\end{equation}

\section{Some Exact Relations}
From their definitions, we can derive the following exact relation about $M_{\alpha \beta, \mu\nu}$ and $N_{\alpha \beta, \mu\nu}$.
From the operator equations
\begin{eqnarray}    \label{eq:C1}
  &&  \sum_{\alpha \beta} f_{\alpha \beta}^{\sigma} \{B_{\alpha \beta}^{\sigma} , d_{\sigma} \} = \{ \{d_{\sigma}, d_{\sigma}^{\dagger} \}, d_{\sigma}\} = 2 d_{\sigma},     \nonumber \\
  &&  \sum_{\alpha \beta} f_{\beta \alpha}^{\sigma \ast} \{D_{\alpha \beta}^{\sigma} , d_{\sigma}^{\dagger} \} = \{ \{d_{\sigma}^{\dagger}, d_{\sigma}\}, d_{\sigma}^{\dagger} \} = 2 d_{\sigma}^{\dagger}, \end{eqnarray}
it is easy to obtain the following relations
\begin{eqnarray}      \label{eq:C2}
&&  \sum_{\alpha \beta} f_{\alpha \beta}^{\sigma} M_{\alpha\beta, \mu\nu} = 2 f_{\mu \nu}^{\sigma},      \nonumber \\
&&  \sum_{\alpha \beta} f_{\beta \alpha }^{\sigma \ast} N_{\alpha\beta, \mu\nu} = 2 f_{\nu \mu}^{\sigma \ast}.
\end{eqnarray}
Similarly, from the operator equations
\begin{eqnarray}    \label{eq:C3}
  &&  \sum_{\alpha \beta} f_{\beta \alpha }^{\sigma \ast} \{B_{\alpha \beta}^{\sigma} , d_{\sigma} \} = \{ \{d_{\sigma}^{\dagger}, d_{\sigma}^{\dagger} \}, d_{\sigma}\} = 0,     \nonumber \\
  &&  \sum_{\alpha \beta} f_{\alpha \beta}^{\sigma } \{D_{\alpha \beta}^{\sigma} , d_{\sigma}^{\dagger} \} = \{ \{d_{\sigma}, d_{\sigma}\}, d_{\sigma}^{\dagger} \} = 0,
\end{eqnarray}
one can get the relations
\begin{eqnarray}      \label{eq:C4}
&&  \sum_{\alpha \beta} f_{\beta \alpha}^{\sigma \ast} M_{\alpha\beta, \mu\nu} = 0,      \nonumber \\
&&  \sum_{\alpha \beta} f_{\alpha \beta }^{\sigma} N_{\alpha\beta, \mu\nu} = 0.
\end{eqnarray}


\begin{thebibliography}{99}
\vspace{0.5cm}
\bibitem{Vollhardt1} W. Metzner and D. Vollhardt, Phys. Rev. Lett {\bf 62}, 324 (1989).

\bibitem{Georges1} A. Georges, G. Kotliar, W. Krauth, M. Rozenberg, Rev. Mod. Phys. {\bf 68}, 13, (1996).

\bibitem{Held1} K. Held, Adv. Phys. {\bf 56}, 829 (2007).

\bibitem{Liebsch1} A. Liebsch, Phys. Rev. B {\bf 70}, 165103 (2004).

\bibitem{Werner1} P. Werner and A. J. Millis, Phys. Rev. Lett. {\bf 99}, 126405 (2007).

\bibitem{Georges2} A. Georges, L. de Medici, and J. Mravlje, Annu. Rev. Condens. Matter Phys. {\bf 4}, 137 (2013).

\bibitem{Kotliar1} G. Kotliar, S. Y. Savrasov, G. P$\acute{a}$lsson, and G. Biroli, Rev. Rev. Lett. {\bf 87}, 186401 (2001).

\bibitem{Hettler1} M. H. Hettler {\it et al.}, Phys. Rev. B. {\bf 58}, R7475 (1998).

\bibitem{Maier1} T. Maier, M. Jarrell, T. Pruschke, and M. H. Hettler, Rev. Mod. Phys. {\bf 77}, 1027 (2005).

\bibitem{Hirsch1} J. E. Hirsch and R. M. Fye, Phys. Rev. Lett. {\bf 56}, 2521 (1986).

\bibitem{Bluemer1} N. Bl$\ddot{u}$mer, Phys. Rev. B {\bf 76}, 205120 (2007).

\bibitem{Werner2} P. Werner {\it et al.}, Phys. Rev. Lett. {\bf 97}, 076405 (2006); P. Werner and A. J. Millis, Phys. Rev. B {\bf 74}, 155107 (2006).

\bibitem{Rubtsov1} A. N. Rubtsov and A. I. Lichtenstein, JETP Lett. {\bf 80}, 61 (2004).

\bibitem{Gull1} E. Gull {\it et al.} Rev. Mod. Phys. {\bf 83}, 349 (2011).

\bibitem{Jarrell1} M. Jarrell and J. Gubernatis, Phys. Rep. {\bf 269}, 133 (1996).

\bibitem{Wilson1} K. G. Wilson, Rev. Mod. Phys. {\bf 47}, 773 (1975).

\bibitem{Bulla1} R. Bulla, Theo A. Costi, and T. Pruschke, Rev. Mod. Phys. {\bf 80}, 395 (2008).

\bibitem{Caffarel1} M. Caffarel and W. Krauth, Phys. Rev. Lett. {\bf 72}, 1545 (1994).

\bibitem{Si1} Q. Si, M. J. Rozenberg, K. Kotliar, and A. E. Ruckenstein,
Phys. Rev. Lett. {\bf 72}, 2761 (1994); M. J. Rozenberg, G. Moeller, and G. Kotliar, Mod. Phys. Lett. B {\bf 8}, 535 (1994).

\bibitem{Liebsch2} A. Liebsch, Phys. Rev. B {\bf 84}, 180505(R) (2011).

\bibitem{Georges3} A. Georges and G. Kotliar, Phys. Rev. B {\bf 45}, 6479 (1992).

\bibitem{Dai1} X. Dai, K. Haule and G. Kotliar, Phys. Rev. B {\bf 72}, 045111 (2005); J. N. Zhuang, Q. M. Liu, Z. Fang, and X. Dai, Chin. Phys. B {\bf 19}, 087104 (2010).

\bibitem{Aryanpour1} K. Aryanpour, M. H. Hettler, and M. Jarrell, Phys. Rev. B {\bf 67}, 085101 (2003).

\bibitem{Bickers1} N. E. Bickers, Rev. Mod. Phys. {\bf 59}, 845 (1987); T. Pruschke, D. L. Cox, and M. Jarrell, Phys. Rev. B {\bf 47}, 3553 (1993); K. Haule, S. Kirchner, J. Kroha, and P. W$\ddot{o}$lflle, Phys. Rev. B {\bf 64}, 155111 (2001).

\bibitem{Zhuang1} J. N. Zhuang, L. Wang, Z. Fang, and X. Dai, Phys. Rev. B {\bf 79} 165114 (2009).

\bibitem{Gros1} C. Gros, Phys. Rev. B {\bf 50}, 7295 (1994).

\bibitem{Jeschke1} H. O. Jeschke and G. Kotliar, Phys. Rev. B {\bf 71}, 085103 (2005); J. X. Zhu, R. C. Albers, and G. Kotliar, Mod. Phys. Lett. B {\bf 20}, 1629 (2006).

\bibitem{Feng1} Q. Feng, Y. Z. Zhang, and H. O. Jeschke, Phys. Rev. B { \bf 79}, 235112 (2009); Q. Feng and P. M. Oppeneer, J. Phys.: Condens. Matter {\bf 24}, 055603 (2012).

\bibitem{Hafermann1} H. Hafermann {\it et al.} Euro. Phys. Lett. {\bf 85}, 27007 (2009).

\bibitem{Julien1} J. P. Julien and R. C. Albers, arXiv:0810.3302.

\bibitem{Granath1} M. Granath and H. U. R. Strand, Phys. Rev. B {\bf 86}, 115111 (2012).

\bibitem{He1} R. Q. He and Z. Y. Lu, Phys. Rev. B {\bf 89}, 085108 (2014); Y. Lu, M. H$\ddot{o}$ppner, O. Gunnarsson, and M. W. Haverkort,  Phys. Rev. B {\bf 90}, 085102 (2014).

\bibitem{Li1} Z. H. Li {\it et al.}, Phys. Rev. Lett. {\bf 109}, 266403 (2012); D. Hou {\it et al.}, Phys. Rev. B {\bf 90}, 045141 (2014).

\bibitem{Hubbard1} J. Hubbard, Proc. Roy. Soc. London {\bf 281}, 401 (1964); F. Gebhard, {\it The Mott Metal-Insulator Transition}
 (Springer-Verlag, Berlin Heidelberg, 1997).

\bibitem{Haley1} S. B. Haley and P. Erd$\ddot{o}$s, Phys. Rev. B {\bf 5}, 1106 (1972).

\bibitem{Haley2} S. B. Haley, Phys. Rev. B {\bf 17}, 337 (1978).

\bibitem{Lacroix1} C. Lacroix, J. Phys. C: Metal Phys. {\bf 11}, 2389 (1981); J. Appl. Phys. {\bf 53}, 2131 (1982).

\bibitem{Luo1} H. G. Luo, J. J. Ying, and S. J. Wang, Phys. Rev. B {\bf 59}, 9710 (1999).

\bibitem{Bulla2} R. Bulla and M. Potthoff, Eur. Phys. J. B {\bf 13}, 257 (2000).

\bibitem{Potthoff1} M. Potthoff, Phys. Rev. B {\bf 64}, 165114 (2001).


\bibitem{Bulla3} R. Bulla, Phys. Rev. Lett. {\bf 83}, 136 (1999).

\bibitem{Tong1} N. H. Tong, S. Q. Shen, and F. C. Pu, Phys. Rev. B {\bf 64}, 235109 (2001).

\end{thebibliography}
\end{document}